\documentclass[superscriptaddress,prd,preprintnumbers,twocolumn,nofootinbib]{revtex4}

\usepackage[utf8]{inputenc}
\usepackage{amsmath,amssymb,amsfonts}
\usepackage{bm, slashed, soul}
\usepackage{graphicx}
\usepackage{hyperref}
\usepackage{multirow}
\usepackage{comment}
\pdfoutput=1 
\usepackage{color}
\definecolor{red}{rgb}{0.9, 0,0}

\def\O{\mathcal{O}}

\newcommand{\be}{\begin{equation}}
\newcommand{\ee}{\end{equation}}
\newcommand{\bea}{\begin{eqnarray}}
\newcommand{\eea}{\end{eqnarray}}

\def\beq#1\eeq{\begin{align}#1\end{align}}

\begin{document}

\title{Oscillons from Pure Natural Inflation}

\author{Jeong-Pyong Hong}
\email{hjp0731@icrr.u-tokyo.ac.jp}
\affiliation{Institute for Cosmic Ray Research, The University of Tokyo, 5-1-5 Kashiwanoha, Kashiwa, Chiba 277-8582, Japan}
\affiliation{Kavli IPMU (WPI), UTIAS, The University of Tokyo, 5-1-5 Kashiwanoha, Kashiwa, Chiba 277-8583, Japan} 
\author{Masahiro Kawasaki}
\email{kawasaki@icrr.u-tokyo.ac.jp}
\affiliation{Institute for Cosmic Ray Research, The University of Tokyo, 5-1-5 Kashiwanoha, Kashiwa, Chiba 277-8582, Japan}
\affiliation{Kavli IPMU (WPI), UTIAS, The University of Tokyo, 5-1-5 Kashiwanoha, Kashiwa, Chiba 277-8583, Japan}
\author{Masahito Yamazaki}
\email{masahito.yamazaki@ipmu.jp}
\affiliation{Kavli IPMU (WPI), UTIAS, The University of Tokyo, 5-1-5 Kashiwanoha, Kashiwa, Chiba 277-8583, Japan} 

\date{\today}

\begin{abstract}
We examine the oscillon formation in a recently proposed inflation model of the {\it pure natural inflation}, where
the inflaton is an axion that couples to a strongly-coupled pure Yang-Mills theory. 
The plateau of the inflaton potential, which is favored by recent observations, 
drives the fragmentation of the inflaton and can produce spatially localized oscillons.
We find that the oscillons are formed for $F\lesssim \O(0.1)M_{\rm pl}$, with $F$ the effective decay constant of the model. 
We also comment on observational implications of the oscillons.
\end{abstract}

\maketitle
\preprint{IPMU 17-0166}


\section{Introduction}

Recently a new inflation model, called the {\it pure natural inflation}, has been introduced in Ref.~\cite{Nomura:2017ehb}.
This is a very simple model where the role of the inflaton is played by an axion, 
which couples to a strongly-coupled pure Yang-Mills theory.
The inflaton here is a pseudo Nambu-Goldstone boson for the shift symmetry of the axion,
and this naturally explains the flatness of the potential \cite{Freese:1990rb,Adams:1992bn}.

The natural question is then if we can test this model in future observations.
In Ref.~\cite{Nomura:2017ehb}, the predictions for the values of the spectral index $n_s$ and the 
tensor-to scalar ratio $r$ has been worked out. The result is in impressive agreement 
with current observational constraints. This is in contrast with the 
natural inflation \cite{Freese:1990rb,Adams:1992bn} with the cosine potential, which has been extensively studied so far but is 
being disfavored by recent results by Planck \cite{Ade:2015lrj} and BICEP/Keck Array \cite{Array:2015xqh}.

The predictions for the values of $r$ and $n_s$ depend crucially on the model parameter $F$,
which plays the role of the effective decay constant of the axion.
When the value of $F$ is large, namely of the order the Planck scale $F\gtrsim \O(m_{\rm pl})$ ($m_{\rm pl} \simeq 1.22 \times 10^{19} \textrm{GeV}$),
the value of $r$ can be of order $r\sim \O(0.1)$, which will be within the reach of future CMB experiment \cite{Creminelli:2015oda}.
When $F$ is smaller and is of order $\lesssim \O(0.1)m_{\rm pl}$,\footnote{See \cite{NY-to-appear} for discussion on theoretical lower bound on the value of $F$.} 
however, $r$ becomes smaller and 
a detection of tensor modes becomes more challenging.
It is therefore an interesting question to see if there is other distinctive 
features of the model, when  $F\lesssim \O(0.1)m_{\rm pl}$.

In this paper we point out that for $F\lesssim \O(0.1)m_{\rm pl}$\footnote{As we will see later, the actual threshold is estimated as a somewhat smaller value:~$F\lesssim\O(0.1)M_{\rm pl}$, with~$M_{\rm pl}=(8\pi)^{-1/2}m_{\rm pl}\simeq2.44\times10^{18}{\rm GeV}$ the reduced Planck mass, which we will mainly use throughout this paper.}
we find interesting new phenomenon absent for larger values of $F$---the generation of 
the spatially-inhomogeneous profile of the inflaton field,
known as oscillon/I-ball~\cite{Bogolyubsky:1976yu,Gleiser:1993pt,Copeland:1995fq,Kasuya:2002zs,Amin:2010dc,Gleiser:2011xj,Amin:2011hj,Lozanov:2017hjm,Salmi:2012ta,Hasegawa:2017iay}.  

It is known that when the scalar field $\phi$ oscillates in the potential that is shallower than quadratic, $\phi$ fragments into quasi-stable lumps, oscillons/I-balls. These can arise in the context of inflation as the fragmentation of the inflaton field after the inflation, whose phenomenon is extensively studied in various models of inflation with shallow potentials~\cite{Amin:2010dc,Amin:2011hj,Takeda:2014qma,Lozanov:2017hjm,Hasegawa:2017iay}. 
Since the small $F$ in our case leads to a flatter potential, as we will see later, the non-linear effect becomes more effective, especially overcoming the damping effect due to the cosmic expansion.
The inflaton fragmentation can have phenomenological implications including the generation of gravitational waves, and spatially localized reheating, which essentially come from the fact the inflaton dominates the universe after the inflation. We will discuss these issues in the later sections.

We first begin with the linear analysis of the growth of the inhomogeneous modes, where in particular we estimate the threshold of $F$ for the instability to overcome the cosmic expansion (Sec.~\ref{sec:dr}),
and next present a full non-linear analysis from numerical lattice simulations (Sec.~\ref{sec:nu}).
We comment on the observational consequences of oscillons in Sec.~\ref{sec:cosmo}.
The final section (Sec.~\ref{sec:conc}) is devoted to conclusions.
We include in App.~\ref{sec:iprf} a review of the I-ball solution.

\vfill

\section{model}\label{sec:model}

In this section let us  summarize salient features of the pure natural inflation \cite{Nomura:2017ehb}.

The potential $V(\phi)$ for the inflaton $\phi$ is given by\footnote{This potential
is the same potential studied in Ref.~\cite{Amin:2011hj}, which reference also studied oscillons for $p<0$.
We emphasize, however, the potential of Eq.~\eqref{eq:potential} here arises dynamically from 
strong-coupling effects of the Yang-Mills gauge field, and predicts a specific sign $p>0$.}
\begin{align}
\label{eq:potential}
V(\phi)=M^4\left[1-\left(1+\left(\frac{\phi}{F}\right)^2\right)^{-p}\right].
\end{align}
Here $M$ determines the overall size of the potential, 
and $F$ is the effective decay constant mentioned above.
The power $p$ is included here to parametrize our ignorance of the 
strongly-coupled gauge theories, and can be determined by improved lattice gauge theory computations
in the near future \cite{NY-to-appear}. 
For the analysis of this paper the precise value of $p\sim \O(1-10)$ is not important,
and in this paper we use the value $p=3$, as suggested by 
holographic computations in the large $N$ limit \cite{Dubovsky:2011tu}.

The potential \eqref{eq:potential} is quadratic $V(\phi)\sim \frac{1}{2} m^2 \phi^2$ 
near the bottom of the potential $\phi\sim 0$, where the value of the inflaton mass 
$m$ is constrained by the observed size of the primordial density perturbations: 
\begin{align}
m=\sqrt{2p}\frac{M^2}{F}\sim 10^{-5} M_{\rm pl},
\label{eq:infm}
\end{align}
where $M_{\rm pl}$ is the reduced Planck mass.
In the opposite limit of $\phi$ large, the potential becomes flat and has a plateau, an attractive feature 
in light of the observed values of $r$ and $n_s$ by Planck \cite{Ade:2015lrj} and BICEP/Keck Array \cite{Array:2015xqh}.
This plateau, which becomes more important as $F$ becomes smaller, will be the origin of the oscillons discussed in this paper.

\section{Linear instability analysis}
\label{sec:dr}

Let us study the growth of the spatial inhomogeneities of the inflaton,
first at the linear level. The question is if
the inhomogeneities as produced by the resonance effect can be strong enough 
to overcome the cosmic expansion.

We divide the inflaton field $\phi$ into the background $\phi_0(t)$ (independent of spatial coordinates $x$) and the fluctuation $\delta\phi(x,t)$:
$\phi(x,t)=\phi_0(t)+ \delta\phi(x,t)$. Their equations of motion are given as follows:
\begin{align}
&\ddot{\phi_0}+3H\dot\phi_0+V'(\phi_0)=0,\\
&\delta\ddot{\phi_k}+3H\delta\dot\phi_k+\left[\frac{k^2}{a^2}+V''(\phi_0)\right]\delta\phi_k=0,\label{eq:fli}
\end{align}
where $\delta\phi_k(t)$ is the Fourier modes for the fluctuation $\delta\phi(x,t)$,
and the dot denotes the derivative with respect to the time $t$.
If $\phi$ is small enough, the background is dominated by harmonic oscillation:
\begin{align}
\phi_0&\simeq\Phi_0\cos (mt),
\end{align}
where $\Phi_0$ is a constant and $m=\sqrt{6}M^2/F$ is the inflaton mass (see Eq.~\eqref{eq:infm}, recall we have chosen $p=3$). 
Then Eq.~(\ref{eq:fli}) for the potential of Eq.~\eqref{eq:potential} is rewritten as
\begin{align}
\delta\ddot{\phi_k}+3H\delta\dot\phi_k+\left[\frac{k^2}{a^2}+m^2-6m^2\left(\frac{\Phi_0}{F}\right)^2\right. \ \ \ \ \ \ \ \ \ \ \ \ \ \ \ \ \ \ \ \nonumber\\
\ \ \ \ \ \left.-6m^2\left(\frac{\Phi_0}{F}\right)^2\cos(2mt)\right]\delta\phi_k\simeq0, \ \ \ 
\end{align}
where we assumed $\Phi_0\lesssim F$ and expanded the potential up to a quartic order.

If we temporarily ignore the cosmic expansion, this becomes the Mathieu equation,
which for $\Phi_0\lesssim F$ has a narrow instability band at 
\begin{align}
\frac{k^2}{a^2}+m^2-6m^2\left(\frac{\Phi_0}{F}\right)^2\simeq m^2,
\end{align}
hence
\begin{align}
\frac k{ma}\simeq\frac{\sqrt{6}\Phi_0}{F}\lesssim \O(1).
\label{eq:isem}
\end{align}
The maximal growth rate of the instability $\mu_{\rm max}$ is given as
\begin{align}
\mu_{\rm max}\simeq\frac{3m}2\left(\frac{\Phi_0}{F}\right)^2,
\label{eq:mumax}
\end{align}
from the analysis of the Mathieu equation~\cite{Kofman:1994rk}.

Of course, we need to take into account the cosmic expansion,
which quickly damps $\phi_0$. 
To overcome this damping effect, we need to have sufficiently large growth rate,
which can be attained by a small $F$, and we 
should obtain an upper bound for such a value of $F$.
The condition for the instability to become non-linear gives 
\begin{align}
1&\sim\frac{\delta\phi}{\phi}
\simeq e^{\mu t}\times\frac{\delta\phi_{\rm initial}}{\phi_{\rm initial}},
\label{eq:mumax1}
\end{align}
where $\mu$ is the growth rate. If we use the observational value for initial fluctuation $\delta\phi_{\rm initial}/\phi_{\rm initial}\sim\mathcal{O}(10^{-5})$, Eq.~(\ref{eq:mumax1}) leads to the following condition:
\begin{align}
\frac{\mu}{H}\sim 10.
\label{eq:muH}
\end{align}
%
%
Since the inflaton at the end of inflation $\phi_{\rm end}$ is not that different from $F$ in this model, we set $\Phi_0\sim\phi_{\rm end}\sim F$. Then, using Eqs.~\eqref{eq:mumax}, \eqref{eq:muH} and also $H\sim m\Phi_0/M_{\rm pl}$, we find that the maximal $F$ for strong resonance is about $\mathcal{O}(0.1)M_{\rm pl}$.

\begin{figure}[htbp]
\centering
  \includegraphics[width=0.9\linewidth]{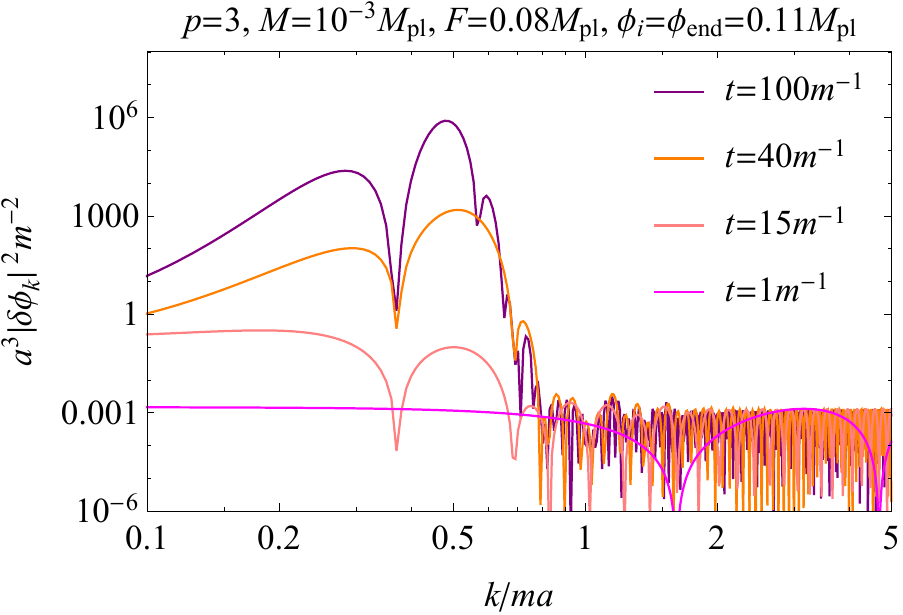}
 \caption{Instability bands for $F=0.08M_{\rm pl},~\phi_i=\phi_{\rm end}\simeq0.11M_{\rm pl}$, which are evaluated at several times. We can see that they are consistent with the rough estimation, Eq.~(\ref{eq:isem}).}
\label{fig:eins}
\end{figure}

That the strong resonance occurs for $F\lesssim0.1M_{\rm pl}$
can also be checked independently by numerically solving Eq.~(\ref{eq:fli}). 
In Fig.~\ref{fig:eins}, we show an example of instability band for $F=0.08M_{\rm pl}$.
Here we set the initial amplitude of the background oscillation as $\phi_i=\phi_{\rm end}\simeq0.11M_{\rm pl}.$\footnote{Strictly speaking, the amplitude of oscillation is larger than $\phi_{\rm end}$, since the velocity is non-zero:~$\dot{\phi}_{\rm end}\neq0$. However, $\phi$ damps to $\phi_{\rm end}$ due to the expansion in a short time scale compared to that of the fragmentation of $\phi$:~$\mathcal{O}(100)m^{-1}$. Since the fragmentation is more efficient for a larger amplitude, our choice of initial condition is rather conservative.}
We note that the instability band is in good agreement with our estimation given in Eq.~(\ref{eq:isem}).
\section{Non-Linear Analysis: Lattice Simulations}
\label{sec:nu}

We expect that the small inhomogeneities produced in the linear analysis 
of the previous section will grow into oscillons---the domination of the quadratic piece of the inflaton potential,
when combined with the shallower higher order corrections, will eventually help to stabilize 
fragmented modes of the inflaton into quasi-stable objects, namely oscillons.

Since this requires us to solve the non-linear dynamics,
we performed lattice simulations by modifying the program {\it LatticeEasy}~\cite{Felder:2007nz}, which is a C++ program designed for simulating the evolution of the scalar fields in an expanding universe with the homogeneous FRW metric. We integrate the equation of motion using the leapfrog method of second order, and approximate the spatial derivatives through the Central-Difference formulas of second order. We confirmed that the numerical error for our lattice~(and time) spacing, which will be presented below, did not affect the order-estimation of the threshold value for the oscillon formation, and the formation time, etc., by performing additional simulations with smaller spacing.
We set the initial amplitude of the background oscillation as $\phi_i=\phi_{\rm end}\simeq0.11M_{\rm pl}$, which is rather a conservative choice as we mentioned in the previous section.
We set the initial scale factor $a$ as unity, and defined the Hubble parameter as
\begin{align}
H=\sqrt{\frac{\langle\rho\rangle}{3M_{\rm pl}^2}},
\end{align}
where $\langle~\rangle$ is the spatial average and $\rho$ denotes the energy density of the inflaton:
\begin{align}
\rho=\frac12\dot{\phi}^2+\frac1{a^2}(\nabla\phi)^2+V(\phi).
\end{align}
In Table~\ref{tab:par}, we present the parameters used in the simulations, including the number of grids, box size, and time step, which are denoted as $N_{\rm grid},L,\Delta t$ respectively.
\begin{table}[t]
\centering
\large
\begin{tabular}{c|ccc}\hline
&{\normalsize $N_{\rm grid}$}&{\normalsize $Lm$}&{\normalsize $m\Delta t$}\\ \hline\hline
{\normalsize1D}&{\normalsize $1024$}&{\normalsize50}&{\normalsize0.05}\\
{\normalsize2D}&{\normalsize$256^2$}&{\normalsize40}&{\normalsize0.11}\\
{\normalsize3D}&{\normalsize$128^3$}&{\normalsize30}&{\normalsize0.14}\\ \hline
  \end{tabular}
 \caption{The parameters used in the simulations including the number of grids, box size, and time step.}
\label{tab:par}
\end{table}
The rescaled program variables are defined as follows:
\begin{align}
\begin{split}
\phi_{\rm pr}&\equiv\phi/\phi_{\rm end},\\
V_{\rm pr}&\equiv V/(m\phi_{\rm end})^{2},\\
t_{\rm pr}&\equiv mt,\\
{\bf x}_{\rm pr}&\equiv m{\bf x}.
\end{split}
\end{align}

\begin{figure}[htbp]
\centering
  \includegraphics[width=0.9\linewidth]{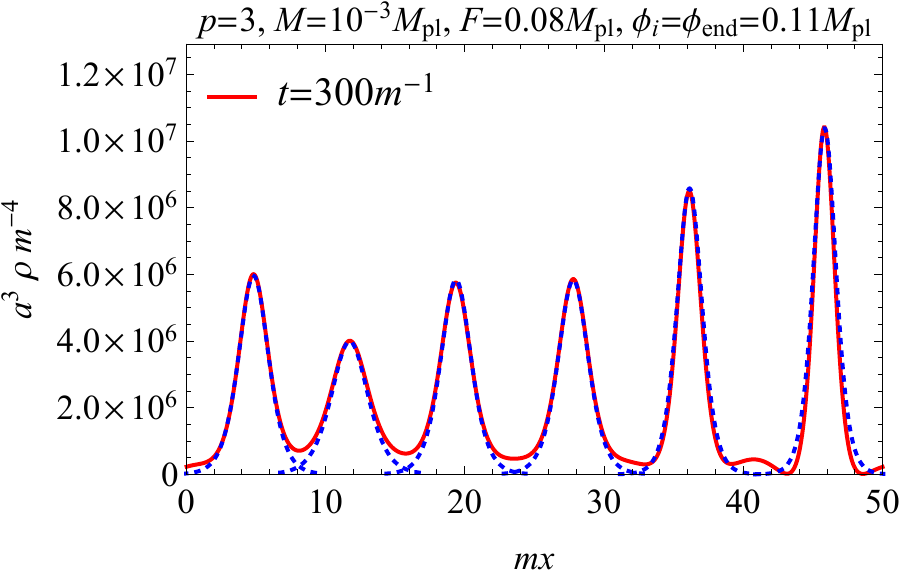}
 \caption{An example of 1D lattice simulations with $F=0.08M_{\rm pl},~\phi_i=\phi_{\rm end}\simeq0.11M_{\rm pl}$, at $t=300m^{-1}$. We set $N_{\rm grid}=1024,~Lm=50$. We plot the comoving energy density normalized by $m^4$, which we fit to the analytic profiles obtained in Sec.~\ref{sec:iprf}. We note that the two are in good agreement.
}
\label{fig:1d}
\end{figure}

In Fig.~\ref{fig:1d}, we illustrate the result of 1D simulations, where we plot energy density $\rho$ after the fragmentation. 
We estimated the formation time as $t_{\rm form}\sim \O(100)m^{-1}$. The reheating temperature without the oscillons is estimated as $T_R\sim10^9~{\rm GeV}$~\cite{Nomura:2017ehb}, whose time scale is much later than the formation time, hence the branching ratio into the other sectors during the formation is negligible.

\begin{figure}[htbp]
\centering
  \includegraphics[width=0.9\linewidth]{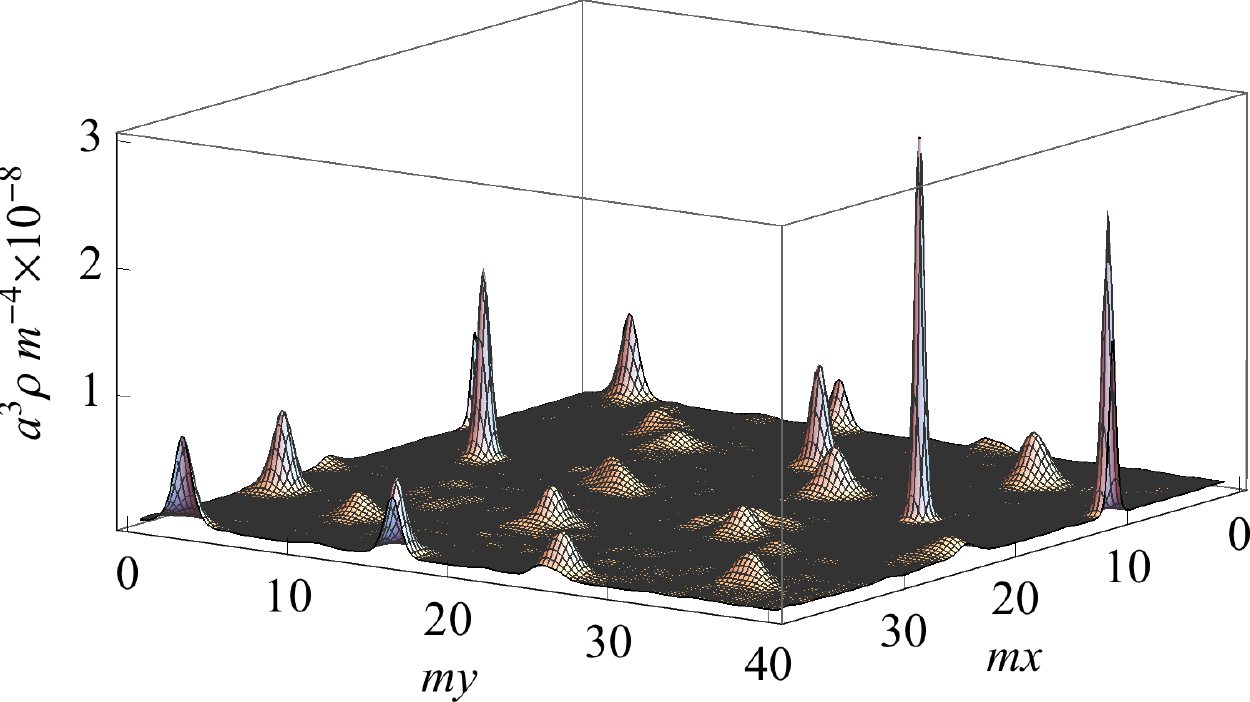}
 \caption{An example of 2D lattice simulation with $F=0.08M_{\rm pl}$ and $\phi_i=\phi_{\rm end}\simeq0.11M_{\rm pl}$, at $t=1000m^{-1}$. We set $N_{\rm grid}=256^2,~Lm=40$. We plot the comoving energy density normalized by $10^8m^4$. }
\label{fig:2d}
\end{figure}
\begin{figure}[htbp]
\centering
  \includegraphics[width=0.8\linewidth]{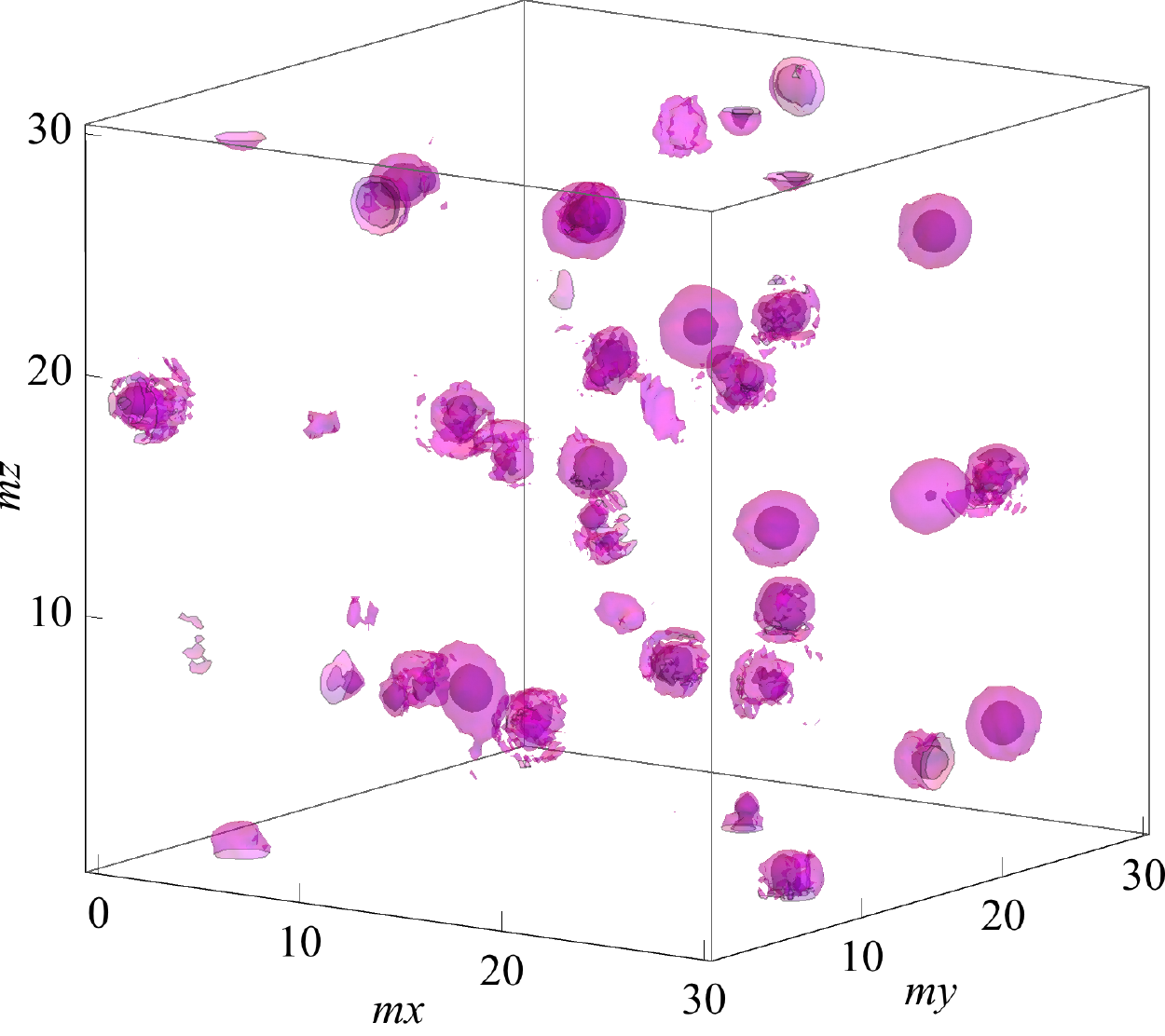}
 \caption{An example of 3D lattice simulation with $F=0.08M_{\rm pl}$ and $\phi_i=\phi_{\rm end}\simeq0.11M_{\rm pl}$, at $t=1000m^{-1}$. We set $N_{\rm grid}=128^3,~Lm=30$. We plot the iso-surfaces of the comoving energy density at $a^3\rho=3.4\times10^{7}m^4$ and $a^3\rho=1.1\times10^{8}m^4$.}
\label{fig:3d}
\end{figure}

We also present the results of 2D and 3D simulations in Figures~\ref{fig:2d} and \ref{fig:3d}, respectively.
In Fig.~\ref{fig:3d}, we plot the iso-surfaces of the energy density at $a^3\rho=3.4\times10^{7}m^4$ and $a^3\rho=1.1\times10^{8}m^4$, where we can indeed identify the localized spherical oscillons.   

It is natural to identify the profile of the resulting oscillons as
I-balls, solutions minimizing energy under the conserving law of the adiabatic invariant (see Sec.~\ref{sec:iprf} for summary).
In particular the energy density as derived from 1D simulations 
is in good agreement with the analytic profile of Eq.~(\ref{eq:ana}),\footnote{Since the analytic profiles are valid for Minkowski spacetime, we compared the numerical results at rather short time, for which the cosmic expansion is negligible.} confirming the 
identification of oscillons with I-balls.
While no analytic solutions of I-balls are known in 2D and 3D, we expect that the similar identification 
holds in these cases as well. We also comment that the discussion on phenomenology in the next section is based on the numerical results, which does not need the analytical approximation on individual profile.

\section{implications}
\label{sec:cosmo}

Having established the presence of oscillons, 
we now comment on their cosmological implications,
in the context of gravitational waves and reheating of the Universe.

\subsection{Gravitational Waves}
Inflaton fragmentation into oscillons can
cause back-reaction to the gravitational modes, and hence 
produce the gravitational waves~(GWs) through the anisotropic stress tensor. 
It is therefore an interesting question to 
compute the frequency $f$ of the such GWs, and see
if it is in the currently observable range (see \cite{Zhou:2013tsa,Antusch:2016con,Antusch:2017flz}) for related recent discussions. 

Since $\mathcal{O}(1{\text -}10)$ I-balls are formed per horizon, a naive expectation is that the frequency corresponds to the Hubble scale:
\begin{align}
f\sim H_{\rm end}&\sim \frac{M^{2}}{M_{\rm pl}}\sim10^{-5}M_{\rm pl} \left(\frac{F}{M_{\rm pl}} \right) \lesssim10^{37}~{\rm Hz},
\end{align}
where we used the relation $F\lesssim \O(M_{\rm pl})$ \cite{Nomura:2017ehb}.
Then, it is redshifted until present, which can be estimated as follows:
\begin{align}
\begin{split}
f_0&=\frac{a_{\rm end}}{a_0}f=\frac{a_{\rm end}}{a_R}\frac{a_R}{a_0}f\\
&\sim 5.5\times10^{-32}\left({\frac{M_{\rm pl}}{\Gamma}}\right)^{1/2}\left(\frac{H_{\rm end}}{\Gamma}\right)^{-2/3}H_{\rm end},\\
&\lesssim10^{8}~{\rm Hz}\times\left(\frac{\Gamma}{10^{-5}M_{\rm pl}}\right)^{1/6},\\
\end{split}
\end{align}
where $\Gamma$ is the decay rate of inflaton~(oscillon) and $a_0$ and $a_R$ represent the scale factors at present and the time of reheating, respectively, and we used $a_{R}/a_0\simeq5.5\times10^{-32}\sqrt{M_{\rm pl}/\Gamma}$ and $a_{R}/a_{\rm end}\simeq(\rho_R/\rho_{\rm end})^{-1/3}\simeq(H_{\rm end}/\Gamma)^{2/3}$.
This is an extremely high frequency compared to the currently observable frequency range~\cite{PhysRevLett.116.061102,Sathyaprakash:2009xs,Moore:2014lga} (this result is consistent with those of Ref.~\cite{Zhou:2013tsa}, which discussed 
inflaton potential \eqref{eq:potential} for $p<0$). Since the result is not highly sensitive to the inflation scale:~$f_0\propto H_{\rm end}^{1/3}$, lowering the frequency by changing the inflation scale will be difficult. Hence we conclude that it is quite generally difficult to produce the GWs in the observable range, through the inflaton fragmentation.

\subsection{Reheating}

Oscillon formation may alter the reheating process after inflation.
If the oscillons are formed,
the universe is reheated by the decay of the spatially-localized oscillons---only 
after the locally generated radiation completely diffuses throughout the space,
we get the spatially-homogeneous radiation era, as is often assumed in 
the standard treatment of reheating.

Let us here define the reheating temperature $T_R$ to be the 
temperature when such spatially-homogeneous radiation 
is first created. We find that there is an upper bound on this temperature.

To derive such a constraint, let us assume that
we indeed have reached an era of homogeneous radiations.
Then the diffusion length $l_d$  of the photons at that time
should be larger than the distance $l_{o{\text -}o}$ between the localized oscillons:
\begin{align}
l_d \gtrsim l_{o{\text -}o}.
\label{eq:ll}
\end{align}
The diffusion length $l_d$ during the time $t\sim H^{-1}$ is estimated to be
\begin{align}
\begin{split}
l_d&\sim l_{\rm mf}\sqrt{N_t}
\sim\sqrt{l_{\rm mf}t}
\sim\frac1{\sqrt{n\sigma H}}
\sim\frac{1}{\sqrt{\alpha^2 TH}}.
\end{split}
\label{eq:ld}
\end{align}
Here $l_{\rm mf}\equiv1/n\sigma$ is the mean free path, with $n\sim T^3$ the number density of the plasma and $\sigma$ the cross section, which we assume as $\sigma\sim\alpha^2/T^2$, with $\alpha$ the structure constant for the radiation.
 $N_t\equiv t/l_{\rm mf}$ is the number of collision during $t$.
The distance $l_{o{\text -}o}$ between oscillons is given by
\begin{align}
\label{eq:loo}
l_{o{\text -}o}&\sim N_o^{-1/3}H^{-1}
\sim N_{o,f}^{-1/3}\left(\frac{H}{H_f}\right)^{1/3}H^{-1},
\end{align}
where $N_o$ denotes the total number of oscillons in the horizon, and the index $f$ refers to the 
quantities at the time of formation of the oscillons. 
Plugging in Eqs.~\eqref{eq:ld} and \eqref{eq:loo} into Eq.~\eqref{eq:ll}
and using $H\sim T_{R}^2/M_{\rm pl}$,
we thus obtain the upper bound on the reheating temperature:  
\begin{align}
T_R\lesssim10^8~{\rm GeV}\alpha^{-6}N_{o,f}^2\left(\frac{H_f}{10^{13}~{\rm GeV}}\right)^2.
\label{eq:rehi}
\end{align}

Now, $H_f$ has the same order of magnitude $H_{\rm end}$, as expected from the 
flatness of the potential (we have also checked this in the numerical simulations).   
Then, using $H_{\rm end}\sim M^2/M_{\rm pl}\sim m F/M_{\rm pl}$ (recall Eq.~\eqref{eq:infm}), Eq.~(\ref{eq:rehi}) reduces to 
\begin{align} 
T_R\lesssim10^8~{\rm GeV}\alpha^{-6}N_{o,f}^2\left(\frac{m_\phi}{10^{13}~{\rm GeV}}\right)^2\left(\frac{F}{M_{\rm pl}}\right)^2.
\end{align}
Hence, for small $F$ and for $\alpha\sim 1$, the upper bound on the temperature for the usual radiation domination becomes severe:~For instance, $F\lesssim0.01M_{\rm pl}$ gives 
a rather stringent constraint $T_R\lesssim 10^4~{\rm GeV}$.\footnote{The decay into the other sectors is negligible during the formation of the oscillons, as mentioned in the previous section. However, they may decay through the self-interactions, before the decay into radiations. The lifetime of the oscillons is not well-understood in the literature. We confirmed that the oscillons are stable up to $t\sim\mathcal{O}(10^5)m^{-1}$ in the 2D simulation with the same setting, and here we simply assume that they are also persistent until the low reheating.}

We emphasize again that the value of the reheating temperature $T_R$ as defined here
is the temperature at the onset of the standard homogeneous radiation era;
the cosmological scenario before this time
can potentially be altered significantly from the standard scenarios, 
due to the the localization of the radiation originating from oscillons.
For example, the temperature at the energy core, which will coincide with the location of oscillons, 
may be higher than predicted by usual perturbative decay rate of the inflaton, which in turn may lead to localized thermal leptogenesis. 
We will discuss these phenomenological consequences in detail in the upcoming paper~\cite{KHPY-to-appear}.

\section{Conclusions}
\label{sec:conc}
We studied inflaton fragmentation in a recently proposed inflation model, called a {\it pure natural inflation}. The small model parameter $F$ gives the flatness to the potential, which drives strong resonance. We found that for $F\lesssim 0.1M_{\rm pl}$, the resonance becomes strong enough to overcome the cosmic expansion, and the inflaton fragments into localized quasi-stable objects called oscillons/I-balls. We confirmed the agreement with the analytical profiles.

We pointed out that the reheating through the oscillon decay may localize the radiation, which must sufficiently diffuse in order to realize the radiation era in the usual sense. We gave an upper bound on the temperature of the beginning of the radiation era by requiring the sufficiently long diffusion length at that temperature.
We found this upper bound for the standard homogeneous radiation domination can be stringent, e.g.,~we obtained $T_R\lesssim 10^4~{\rm GeV}$ for $F\lesssim0.01M_{\rm pl}$. The localization of the radiation can be important since the high temperature phenomena such as thermal leptogenesis may occur in a localized manner, which we will pursue in the future work.


\section*{Acknowledgements}
We would like to thank Fuminori Hasegawa  and Yasunori Nomura for helpful discussion.
This work is supported by WPI Initiative, MEXT, Japan.
MK is supported in part by MEXT KAKENHI Grant No.\ 15H05889, and JSPS KAKENHI Grant Nos.\ 17K05434 and 17H01131.
MY is supported in part by JSPS KAKENHI Grant No.\ 15K17634, and by JSPS-NRF Joint Research Project.

\begin{appendix}
\section{I-ball profiles}
\label{sec:iprf}

In this Appendix we review the I-ball solutions following Ref.~\cite{Kasuya:2002zs}.

When the inflaton energy damps due to the cosmic expansion, the harmonic oscillation
around the minimum $\phi=0$ of the potential dominates the inflaton dynamics. The area enclosed by the periodic trajectory in the phase space is then conserved to a good approximation, which quantity we call the adiabatic invariant $I$:
\begin{align}
I\equiv\frac1{2\omega}\int d^3x\, \overline{\dot{\phi^2}},
\end{align}
where the bar denotes a time-average over a period of the periodic motion, 
and $\omega$ is the frequency of the harmonic oscillation,
which in practice is approximately given by the inflaton mass $m$: $\omega\sim m$.

I-balls are defined as the solutions that minimize the energy (averaged over one period) for a given $I$~\cite{Kasuya:2002zs}. 
To find them, we use the Lagrange multiplier method and minimize the following quantity:
\begin{align}
E_\omega&\equiv\overline{E}+\tilde{\omega}\left(I-\frac1{2\omega}\int d^3x\, \overline{\dot{\phi^2}}\right)\\
&=\int d^3x\left[\left(1-\frac{\tilde{\omega}}{\omega}\right)\frac12\overline{\dot{\phi^2}}+\frac12\overline{(\nabla\phi)^2}+\overline{V(\phi)}\right]\\
&\simeq\int d^3x\left[\left(1-\frac{\tilde{\omega}}{\omega}\right)\frac14\omega^2\Phi^2+\frac14(\nabla\Phi)^2\right.\nonumber\\
&\ \ \ \ \ \ \ \ \ \ \ \ \ \ \ \ \ \ \ \ \ \ \ \ \ \ \left.+\frac14\omega^2\Phi^2-\frac{6M^4}{F^4}\frac38\Phi^4\right].
\label{eq:eome}
\end{align}
Here $\tilde\omega$ is the Lagrange multiplier and $\Phi$ denotes the amplitude of the oscillation
({\it i.e.} $\phi(x,t)=\Phi(x) \cos (\omega t)$), and we used the time-averaged quantities, which are given as
\begin{align}
\overline{\phi^2}&\simeq\frac12\Phi^2,\\
\overline{\dot{\phi^2}}&\simeq\frac12\omega^2\Phi^2,\\
\overline{V(\phi)}&\simeq\frac{3M^4}{F^2}\frac12\Phi^2-\frac{6M^4}{F^4}\frac38\Phi^4 \nonumber\\
&\simeq\frac14\omega^2\Phi^2-\frac{6M^4}{F^4}\frac38\Phi^4.
\end{align}
Note that we approximated the potential by keeping terms up to quartic order in the field $\phi$.
By varying Eq.~(\ref{eq:eome}) with respect to $\Phi$, we obtain the following equation for the amplitude
\beq
\Delta\Phi-\left(2-\frac{\tilde{\omega}}{\omega}\right)\omega^2\Phi+\frac{18M^4}{F^4}\Phi^3=0,
\eeq
which defines I-balls.

For 1+1 dimension we find the following analytic solution for 1+1 dimension:
\beq
\Phi(x)=\Phi(0)\, {\rm sech}\left[\left(\frac{\sqrt{3}M}{F}\right)^2\Phi(0)x\right],
\label{eq:ana}
\eeq
where we used the expression for $\omega\sim m$ given in Eq.~\eqref{eq:infm} (with $p=3$),
and $\tilde{\omega}$ is traded for a constant $\Phi(0)$.
This profile is consistent with that obtained by solving the equation of motion in small amplitude approximation~\cite{trove.nla.gov.au/work/8361324,Fodor:2008es}, whose method is especially useful when the potential is asymmetric and the anharmonic correction becomes important~\cite{Antusch:2017flz,Hasegawa:2017iay}.
In Sec.~\ref{sec:nu}, we compare this analytic profile to that obtained from 1D lattice simulations,
which gives a strong evidence that oscillons are identified with I-balls.

\end{appendix}
\bibliography{references}

\end{document}